\documentclass[journal]{IEEEtran}
\usepackage{graphicx}
\usepackage[margin=1in]{geometry}
\usepackage{amssymb,amsfonts,amsmath,amsthm,amscd,dsfont,mathrsfs,mathtools}
\usepackage{graphicx,float,psfrag,pstcol,epsfig,color,subfigure}
\usepackage{psfrag}
\usepackage{pstcol}
\usepackage{import}
\usepackage{setspace}
\usepackage{url}

\newcommand{\ux}{x^n}
\newcommand{\uy}{y^n}
\newcommand{\uX}{X^n}
\newcommand{\uY}{Y^n}
\newcommand{\cY}{{\cal Y}}
\newcommand{\cX}{{\cal X}}
\newcommand{\K}{\psi}
\newcommand{\bX}{{\bf X}}
\newcommand{\bY}{{\bf Y}}
\newcommand{\bN}{{\bf N}}
\newcommand{\bA}{{\bf A}}
\newcommand{\bR}{{\bf R}}
\newcommand{\bH}{{\bf H}}

\theoremstyle{definition}
\theoremstyle{plain}
\theoremstyle{plain}
\theoremstyle{plain}
\theoremstyle{plain}
\theoremstyle{plain}
\theoremstyle{plain}
\theoremstyle{remark}
\newtheorem{remark}{Remark}

\begin{document}

\title{Communication Theoretic Data Analytics}

\author{Kwang-Cheng Chen, Shao-Lun Huang, Lizhong Zheng, H. Vincent Poor 
\thanks{This research was supported in part by the US Air Force under the grant AOARD-14-4053, the US NSF grant CCF-1216476, the U.S. Army Research Office under MURI Grant W911NF-11-1-0036, and the U. S. National Science Foundation under Grant CCF-1420575.}
\thanks{K.C. Chen and S.-L. Huang are with the Graduate Institute of Communication Engineering, National Taiwan University, ckc@ntu.edu.tw}
\thanks{S.-L. Huang and Lizhong Zheng are with the Department of Electrical Engineering and Computer Science, Massachusetts Institute of Technology, shaolin@mit.edu and lizhong@mit.edu}
\thanks{H.V. Poor is with the Department of Electrical Engineering, Princeton University, poor@princeton.edu }
}


\maketitle

\begin{abstract}
Widespread use of the Internet and social networks invokes the generation of big data, which is proving to be useful in a number of applications. To deal with explosively growing amounts of data, data analytics has emerged as a critical technology related to computing, signal processing, and information networking. In this paper, a formalism is considered in which data is modeled as a generalized social network and communication theory and information theory are thereby extended to data analytics. First, the creation of an equalizer to optimize information transfer between two data variables is considered, and financial data is used to demonstrate the advantages. Then, an information coupling approach based on information geometry is applied for dimensionality reduction, with a pattern recognition example to illustrate the effectiveness. These initial trials suggest the potential of communication theoretic data analytics for a wide range of applications.
\end{abstract}


\IEEEpeerreviewmaketitle

\section{Introduction}

With the booming of Internet and mobile communications, (big) data analytics has emerged as a critical technology, adopting techniques such as machine learning, graphical models, etc. to mine desirable information for a wide array of information communication technology (ICT) applications~\cite{1}\cite{2}\cite{16}\cite{17}\cite{Jensen}\cite{Koller}. Data mining or knowledge discovery in databases (KDD) has been used as a synonym for analytics on computer generated data. To achieve the purpose of data analytics, there are several major steps: (i) based on the selection of data set(s), pre-processing of the data for effective or easy computation, (ii) processing of data or data mining, likely adopting techniques from statistical inference and artificial intelligence, and (iii) post-processing to appropriately interpret results of data analytics as knowledge. Knowledge discovery aims at either verification of user hypotheses or prediction of future patterns from data/observations. Statistical methodologies deal with uncertain or nondeterministic reasoning and thus models, and are the focus of this paper. Machine learning and artificial intelligence are useful to analyze data, e.g.~\cite{2}\cite{16}\cite{18}, typically via regression and/or classification. With advances in supervised and unsupervised learning, inferring the structure of knowledge, such as inferring Bayesian network structure from data, is one of the most useful information technologies~\cite{19}. In recent decades, considerable research effort has been devoted to various aspects of data mining and data analysis, but effective data analytics are still needed to address the explosively growing amount of data resulting from Internet, mobile, and social networks. 

A core technological direction in data analytics lies in processing high-dimensional data to obtain low-dimensional information via computational reduction algorithms, namely by nonlinear approaches~\cite{20}\cite{21}, compressive sensing~\cite{22}, or tensor geometric analysis~\cite{23}. In spite of remarkable achievements, with the exponential growth in data volume, it is very desirable to develop more effective approaches to deal with existing challenges including effective algorithms of scalable computation, complexity and data size, outlier detection and prediction, etc. Furthermore, modern wireless communication systems and networks to support mobile Internet and Internet of Things (IoT) applications require effective transport of information, while appropriate data analytics enable communication spectral efficiency via proper context-aware computation~\cite{24}. The technological challenge for data analytics due to very large numbers of devices and data volume remains on the list of the most necessary avenues of inquiry. At this time, state-of-the-art data analytics primarily deal with data processing through computational models and techniques, such as deep learning~\cite{25}. There lack efforts to examine the mechanism of data generation~\cite{26} and subsequent relationships among data, which motivates the investigation of data analytics by leveraging communication theory and information theory in this paper. 

As indicated in~\cite{3} and other works, relationships among data can be viewed as a type of generalized social network. The data variables can be treated as nodes in a network and their corresponding relationships can be considered to be links governed by random variables (or random processes by considering a time index). Such scenarios are particularly relevant for today's interactive Internet data from/related to social networks, social media, collective behavior from crowds, and sensed data collected from sensors in cyber-physical systems (CPS) or Internet of Things. Therefore, with the aid of this generalized social network concept, we propose a new communication theoretic methodology of information-centric processing for (big) data analytics in this paper. Furthermore, by generalizing the scenario from a communication link into a communication network, we may use ideas from network information theory and information geometry to develop a novel technology known as information coupling~\cite{HuangZheng}, which suggests a new information-centric approach to extraction of low-dimensional information from high-dimensional data based on information transfer. These technological opportunities describe a complete communication theoretic view of data analytics. 

The rest of this paper is organized as follows. Section II presents our modeling of data into a generalized social network and its resemblance to typical communication system models. Section III describes the setting of our proposed communication theoretic data analytics to more effectively process the data, using financial data to illustrate the processing methodology with comparisons to well-known techniques. Related literature is reviewed to better explain our proposed methodology. Section IV briefly introduces the rationale from information geometry and the principle of information coupling to realize dimensionality reduction, with an image pattern recognition example to show the effectiveness of this new idea based on network information theory. Finally, we make concluding remarks in Section V, with suggested open problems to fully understand and to most effectively revisit data mining and knowledge discovery in (big) data analytics. In addition to its potential for creating new methods for data analytics, this new application scenario also creates a new dimension for communication and information theory.

\section{Social Network Modeling of Data}

As noted in~\cite{3}, entities (e.g. data) with relationships can be viewed as social networks, and thus social network analysis and statistical communication theory share commonalities in many cases. For data analytics, it is common to face a situation in which we have two related variables, say $X$ and $Y$. When there exists uncertainty in observing these two variables, it is common to model these two variables as random variables. If a time index is involved, say the variables are observed or sampled in sequence, these two variables are actually two random processes. Consequently, each sequence of data drawn from a variable is actually a sample path (i.e. sampled data) of the random process. An intuitive way of examining the relationship between the two processes is to compute the correlation coefficient between these two sampled data sequences. 

For big data problems in an Internet setting, we are often facing a substantial number of variables up to thousands or even millions in order, and therefore must rely on machine learning to handle such scenarios to predict or otherwise to infer from data. One of the key problems is to identify low-dimensional information from high-dimensional data, as a key issue of knowledge discovery. Recently, another vision, known as \emph{small data}, has emerged to more precisely deal with variables of data on a human scale~\cite{10}. Therefore, in data analytics, whether addressing big data or small data, effective and precise inference from data is always the fundamental challenge. An approach different from machine learning arises by extracting embedded information from data. More precisely, for example, we may identify the information flowing from variable $X$ to variable $Y$ just as in a typical point-to-point digital communication system.

Unfortunately, real world problems are much more complicated than a single communication link, and there are many more variables involved. Figure 1 depicts the social network structure of a large data set through realization of graphical models and Bayesian networks, while each node (i.e. variable) represents a data variable (actually a vector of data) and each link represents the relationship and causality between two data variables. Such relationships between two data variables usually exhibit uncertainty due to the nature of data or imperfect observations, and thus require probabilistic modeling. Even more challenging, such causal relationship among large numbers of variables may not be known, and thus a challenge is to determine or to learn the \emph{knowledge discovery} structure~\cite{Koller}\cite{Koivisto}\cite{28}.

\begin{figure}
\centering 
\includegraphics{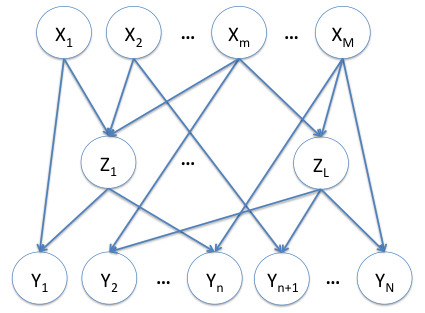} 
\caption{Graphical model of network variables for a large data set.}
\label{fig:figure_1}
\end{figure}

The social network analysis of data can be performed in different ways, such as using graphical models with machine learning techniques~\cite{2}\cite{3}. However, as noted, such widely applied methodologies focus on data processing and inference, rather than considering information transfer. Communications can be generally viewed as transmission of information from one location to another location as illustrated in Figure 2(a). We may further use signal flow of random variables to abstractly portray such a system as in Figure 2(b). The channel as a link between random variables $X$ and $Y$, can be characterized by the condition probability distribution $f(y|x)$. When a channel involves multiple intermediate variables relating $X$ and $Y$, this results in receive diversity as shown in Figure 2(c).

\begin{figure}
\centering 
\includegraphics[width=\columnwidth]{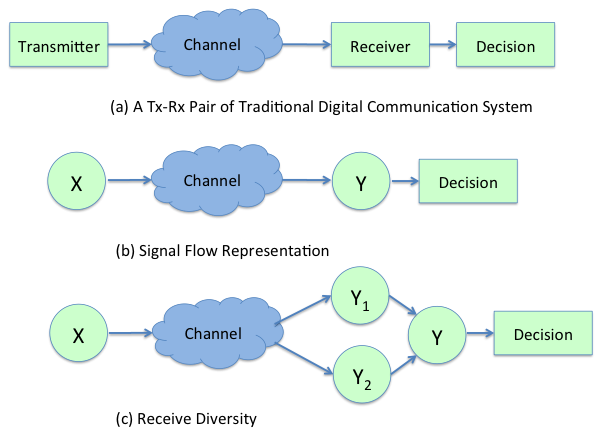} 
\caption{Communication theory in signal flow or graphical model to show causal relationship in data variables.}
\label{fig:figure_2}
\end{figure}

More advanced communication theory, namely multiple access communication, has been developed in recent decades and may be useful for Internet data analytics. Multiuser detection (MUD), though commonly connected with code division multiple access (CDMA), generally represents situations in which multiple user signals (no need to be orthogonal) are simultaneously transmitted then detected over the same frequency band~\cite{4}. In such situations, the signal model can be described as
\begin{align*}
\bY=(\bA \bR) \bX+ \bN
\end{align*}
where $\bX$ is the transmitted signal vector; $\bY$ is the received signal vector embedded in noise $\bN$; $\bR$ denotes the correlation matrix among signals used by transmitter-receiver pairs and $\bA$ is a diagonal matrix containing received amplitudes. The non-diagonal part of $\bA \bR$ results in multiple-access interference (MAI). Similarly, a multiple antenna (MIMO) signal model can be described mathematically as
\begin{align*}
\bY= \bH \bX+ \bN
\end{align*}
where $\bH$ is the channel gain matrix~\cite{5}. From the similarity in mathematical structure of MUD and MIMO systems, a duality in receiver structures can be also observed. This is fundamentally the same information flow structure as in data analytics, and so multiuser communication theory provides a new tool to comprehend information flow in general social networking models. In~\cite{3}, recommender systems are illustrated as an example for this potential. 

\begin{figure}
\centering 
\includegraphics[width=\columnwidth]{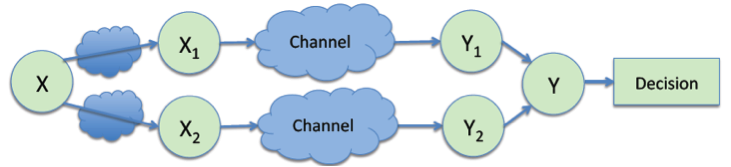} 
\caption{Distributed detection for sensor networks.}
\label{fig:figure_3}
\end{figure}

In this paper, we will delineate the connection between data analytics and social network analysis, considering a link in a generalized social network to be equivalent to a communication link. Consequently, we can leverage the knowledge of communication theory to investigate data analytics, a process we term \emph{communication theoretic data analytics}. Processing on graphs to extract motif information aims at alleviating the complexity of data analytics~\cite{Sandryhalia}, and  bears somewhat the same spirit as the communication theoretic data analytics. Section III will introduce the optimal receiver to tackle nonlinear distortion by using an equalizer, and thereby to optimize information transfer for more effective data analytics. A further interesting communication model is the sensor network illustrated in Figure~\ref{fig:figure_3}, where an information source is detected by multiple sensors that send their measurements to a fusion center for decisions~\cite{6}. The number of sensors might be large but the actual information source might be simple. Directly processing sensor data might be a big data problem but a single source may induce simplification by considering information transfer, which alternatively suggests an information theoretic formulation of information coupling toward critical dimension reduction in data analytics, which will be introduced in Section IV.

\section{Communication Theoretic Implementation of Data Analytics and Applications}

Using the communication theoretic setup, we will demonstrate how to deal with practical data analytics. To infer useful results from big data, we will be able to acquire some knowledge, say the conditional probability structure $f(y_n |x_m)$ in a general social network modeling of big data as in Figure 1. Through communication theory, we may treat $X_m$ as an information transmitter and $Y_n$ as the information receiver. In our setting, $f(y_n|x_m)$ represents the communication channel, which can be corrupted by noise due to imperfect sampling, noisy observation, and possible interference from unknown variables or outliers. This point-to-point scenario is well studied in communication theory. 

A common and most fundamental scenario in (big) data analytics is to find the relationship between two variables, $X$ and $Y$, while each variable represents a set or a series of data. Let us borrow the concept of Bayesian networks or graphical models in machine learning. We may subsequently infer the realization $y$ of $Y$ based on the observation of the realization $x$ of $X$. Suppose there exists a causal relationship $X \rightarrow Y$, with uncertainty (or embedded in an unknown mechanism allowing information transfer). As noted above, this causal relationship can be represented by the conditional or transition probability $f(y|x)$, which can be considered as a communication channel. Direct computation and inference therefore proceed based on this structure. If we intend to infer the knowledge structure such as this simple causal relationship, machine learning on (big) data becomes a principal problem~\cite{2}\cite{Jensen}\cite{Koller}\cite{18}. 

\subsection{Equalizer to Optimize Information Transfer}

Again, our view of this simple causal relationship considers information transfer from $X$ to $Y$ as a communication channel. In this context, if we want to determine the causal relationship of data due to information transfer, according to communication theory, we should establish an equalizer for the channel to better receive such information. The equalizer in data analytics is most appropriately of the form of an adaptive equalizer, with earlier observed data used for the purpose of training to obtain the weighting coefficients. The most challenging task in machine learning is knowledge discovery, i.e., identifying the causal relationship among variables. The communication theoretic approach supplies a new angle from which to examine this task, and its information theoretic insight will be investigated in Section IV via information coupling. 

\begin{figure}
\centering 
\includegraphics[width=\columnwidth]{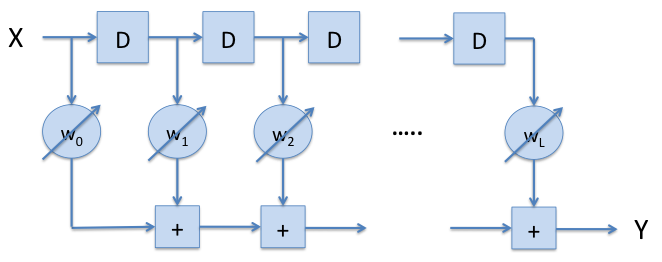} 
\caption{Equalizer structure to extract the causal relationship between two variables, from $X$ to $Y$.}
\label{fig:figure_4}
\end{figure}

Well known in communication theory, an equalizer serves as a part of an optimal receiver to detect a transmitted signal from a noisy channel distorted by nonlinearities and other impairments. In order to infer from another data variable or time series, we may take advantage of the same concept to process the data. Proper implementation of an equalizer could enhance the causal relationship between data variables and time series, and thus allow better judgement or utilization of the causal relationship. This is one core issue in knowledge discovery of data analytics. In big data analytics, this knowledge problem has a very large number of variables. As in Figure 1, we focus on the problem of identifying the causal relationships between the set of variables $X_1,X_2, \cdots ,X_M$ and the set of variables $Y_1,Y_2, \cdots ,Y_N$, specified by appropriate weights. This is identical to the following multiple access communication problem:
\begin{align} \label{equation_1}
(X_1,X_2, \ldots ,X_M ) [h_{ij} ]_{M \times N}=(Y_1,Y_2, \ldots ,Y_N )^T
\end{align}
where $[h_{ij} ] = {\bf H}$ is analogous to the channel matrix. This knowledge discovery problem in data analytics is thus equivalent to a blind channel estimation/identification problem in multiple access communication. Since a feedback channel may not exist in general, this is a blind problem. However, for online processing, equivalent feedback might not be impossible, which we leave as an open issue. We start from some simple cases to illustrate this idea.

\begin{itemize}
\item {\bf Information diversity}: A variable $X$ may influence a number of variables, say $Y_1,Y_2, \cdots ,Y_N$, which is a form of information diversity. To identify a causal relationship between $X$ and $Y_n$, this is precisely a multi-channel estimation problem such as arises in wireless communication. Since feedback in causal data relationships is generally impossible, such a class of problems falls under the category of blind multi-channel estimation/identification~\cite{11}\cite{Yu}.

\item {\bf Information fusion}: Another class to consider is the causal relationship from many data variables to influence a single data variable, say $X_1,X_2, \cdots ,X_M$ to $Y$. This corresponds to multi-input-single-output channel estimation or identification, which is a rather overlooked subject. However, another similar problem, source separation has been well studied. 
\end{itemize}

\subsection{Applications to Inference on Financial Time Series}

\begin{figure*}
\centering 
\includegraphics[width=2\columnwidth]{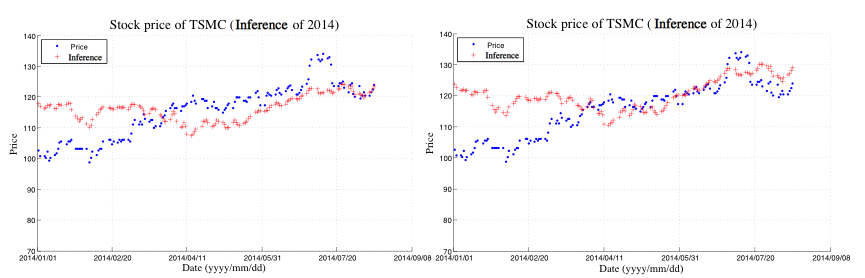} 
\caption{Prediction of TSMC stock price by (a) multivariate linear regression, and (b) multivariate Bayesian estimation, where blue dots denote actual prices and red crosses denote prices inferred from the exchange rate and NASDAQ index.}
\label{fig:figure_6}
\end{figure*}

A useful way to demonstrate our analytical methodology is to consider financial time series data, which has been well studied in the literature. The purpose is to demonstrate the prediction of stock prices from other factors. In this example, we are trying to predict the stock price of Taiwan Semiconductor Manufacturing Corp. (TSMC), which is the world's largest contract chip fabrication company. To demonstrate information transfer and thus communication theoretic data analytics, we consider two factors, which appear to be somewhat indirectly related to stock prices in Taiwan but are potentially influential: the exchange rate between US dollars (USD) and New Taiwan Dollars (NTD, the local currency in Taiwan), and the NASDAQ index which primarily consists of high-tech stocks in the US. 

Let the time series of the exchange rate between USD and NTD be $X_1 [n]$, $n=0,1, \ldots$ and the time series of the NASDAQ index be $X_2 [n]$, $n=0,1, \ldots$. The time series for the stock price of TSMC is $Y[n]$, $n=0,1, \ldots$. Now, both $X_1 [n]$ and $X_2 [n]$ may influence $Y[n]$. This classical problem has been typically handled by multivariate time series analysis, which serves as a benchmark without introducing more advanced techniques. 

Now we treat this \emph{information fusion} problem as a two-channel information transfer (and thus communication) problem: $X_1 [n] \rightarrow Y[n]$ and $X_2 [n] \rightarrow Y[n]$. To proceed, we establish an equalizer to filter the data in each channel. The equalizer is typically of the tap-delay-line type, while the time unit is one day since our data uses the closing rate/index/price. The length of this adaptive equalizer is $L$ and the corresponding weighting coefficients are determined via training. We treat 2009-2013 as the training period and then infer 2014 data in an online way. The order of the adaptive equalizer and the weighting coefficients are learned during the training period, and they are kept and used during inference. Our numerical experiment shows that

\begin{figure}
\centering 
\includegraphics[width=\columnwidth]{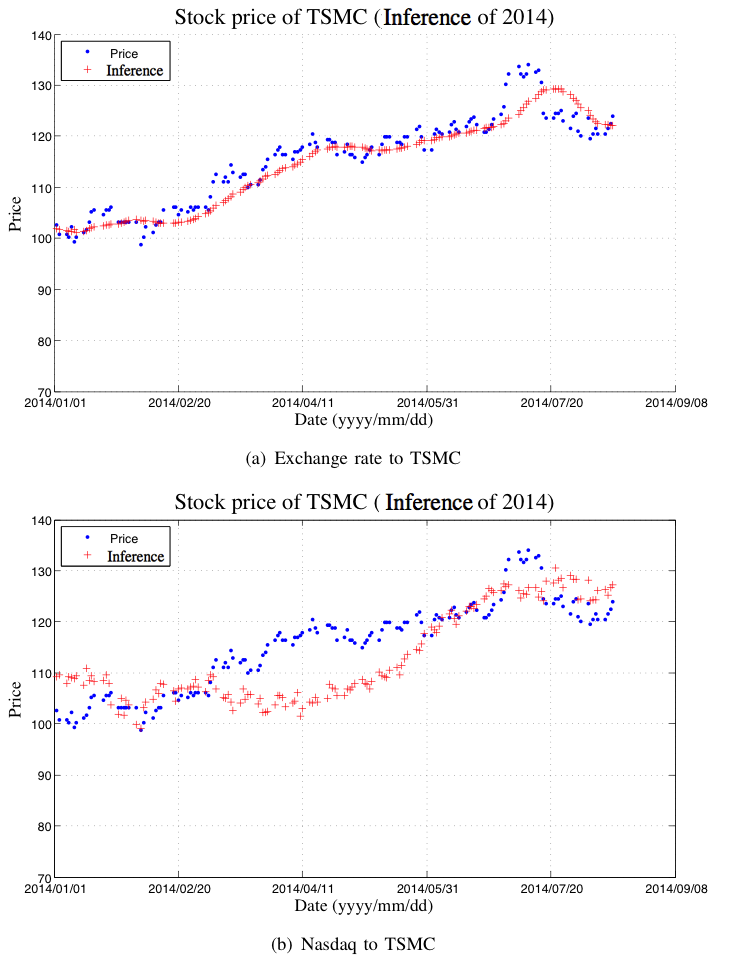} 
\caption{Communication theoretic data analytics using an equalizer to optimize information transfer from (a) the exchange rate, and (b) the NASDAQ index to the TSMC stock price.}
\label{fig:figure_7-2}
\end{figure}

\begin{figure}
\centering 
\includegraphics[width=\columnwidth]{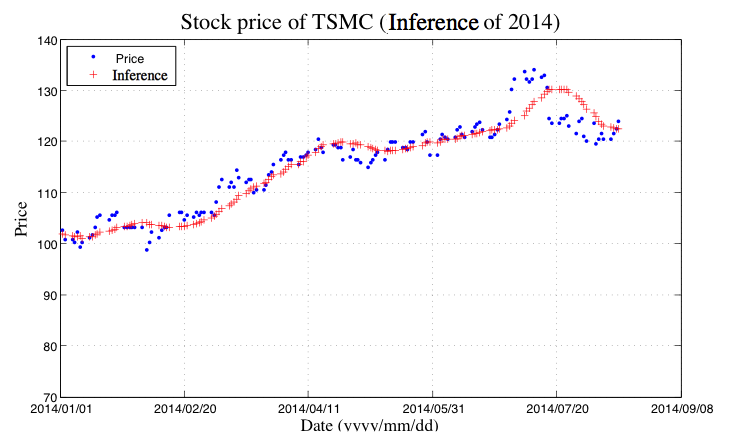} 
\caption{Inference via MRC information combining two equalized data processing channels, where the length of the equalizer for the exchange rate is 16 (days) and the length of the equalizer for the NASDAQ index is 61 (around 3 months).}
\label{fig:figure_8-2}
\end{figure}

\begin{itemize}
\item Each individual factor (exchange rate or NASDAQ index) is surprisingly useful but each alone is not good enough to infer the TSMC stock price. Furthermore, via classical methods such as linear least squares or Bayesian estimation~\cite{29}, the exchange rate appears to be a much less predictive factor than the NASDAQ index, which is to be expected since the NASDAQ index to a certain extent can represent high-tech stocks including TSMC. 
\item It is expected that multivariate statistical analysis would help in this case. We adopt multivariate linear regression~\cite{30} and multivariate Bayesian estimation~\cite{29} as benchmark techniques. The inference of TSMC stock prices from the exchange rate and NASDAQ index is shown in Figure~\ref{fig:figure_6}. The mean square errors for both techniques perform similarly to the results using only the NASDAQ index. 
\item We implement the tap-delay-line equalizer structure of Figure~\ref{fig:figure_4} to optimize information transfer. Based on the mean square error (MSE) criterion, we search for the best length equalizer and corresponding weighting coefficients, which are then used for inference. During the inference period of 2014, the length remains the value from training but the weighting coefficients are updated online. In Figure~\ref{fig:figure_7-2}, we surprisingly observe excellent performance of inference using the exchange rate, as effective as the NASDAQ index. This result illustrates different insights into the correlation of data from the traditional approaches, since these approaches suggest that the NASDAQ index can describe the TSMC stock better. Therefore, our result demonstrates the potential of this communication theoretic data processing methodology and the potential of considering information transfer. Thus, the potential of information-centric data processing over conventional machine learning is worth further study.
\item Similar to diversity combining in digital communication systems, information combining in communication theoretic data analytics potentially further improves the performance of inference. Maximal ratio combining (MRC) is well known to be optimal for diversity combining based on signal-to-interference-plus-noise ratio (SINR). Using a similar concept as equation~\eqref{A5}, we develop MRC information combining to weight equalized channels inversely proportional to the MSE in the training period. Such weights in MRC information combining can be updated online. Figure~\ref{fig:figure_8-2} depicts the prediction results (red crosses) and true values (blue dots). We calculate the mean square error and find even better performance than using only the exchange rate. The MSE can be lower than those of multivariate linear regression and multivariate Bayesian estimation.
\end{itemize}

In Appendix A, we develop the following rules of thumb for communication theoretic data analytics, while subject to further enhancements.

{\noindent Problem: To infer $Y$ based on $X_1,X_2, \ldots ,X_N$.}\\
{\noindent Procedure:}
\begin{itemize}
\item [(1)] Use an equalizer (i.e. optimal receiver) implementation to identify causal relationships among data variables, $X_1 \rightarrow Y, X_2 \rightarrow Y , \ldots ,X_N \rightarrow Y$, with the corresponding MSEs according to the training dataset(s).
\item [(2)] Select $N_c$ data variables to transfer sufficient information (or sufficiently small MSE errors in training) to identify the structure of knowledge by keeping the length and coefficients of the equalizer, or by online update of coefficients.
\item [(3)] Conduct MRC information fusion of these $N_c$ data variables as in Fig. 12  to infer.
\end{itemize}

\begin{remark}
The conjecture that this communication theoretic data analytics approach delivers more desirable performance comes from optimizing information transfer and avoiding cross-interference among data variables (similar to multiple access interference in multiuser communication), while existing multivariate statistical analysis or statistical learning multiplexes all data variables together to result in multiple access interference in data analytics. Furthermore, for each data variable, a selected equalizer length with coefficients is used, then information is combined with other data variables, to allow better matching to extract information in this communication theoretic data analytics approach. Each equalizer is designated to match a specific data variable, while multivariate analysis usually deals with a common fixed depth of observed data in processing for all data variables. More observations may not bring in more relevant information but rather additional noise/interference. Although recent research suggests a duality between time series and network graphs~\cite{31}\cite{32}, information-centric processing of data suggested by communication theory supplies a unique and generally applicable view of inference, even though its extension to more complicated network graphical relationship of data is still open. Note that during the training period, the computational complexity is high, however, the computation load is rather minor in the inference stage.
\end{remark}

\begin{remark}
Applying information theoretic data analytics such as mutual information and information divergence beyond correlation have been proposed in the literature, e.g.~\cite{33}\cite{34}. However, such efforts have not systematically applied information-centric processing techniques based on communication systems as suggested in this paper. In the mean time, though we have illustrated only one-hop network graphical inference, these methods may be applied further for data cleaning, data filtering, identification of important data variables for inference, and identification of causal relationships among data variables to support knowledge discovery in data analytics. To fuse heterogeneous information, fuzzy logic~\cite{33} or Markov logic~\cite{34} is usually employed. In the proposed approach, information combining of different-depth information transfer alternatively serves the purpose in an effective way, while time series data mining typically considers similarity measured in terms of distance~\cite{Esling}.
\end{remark}

At this point, we cannot conclude that communication theoretic data analytics are better than multivariate analysis and other machine learning techniques, as many advanced techniques such as Kalman filtering~\cite{Harvey}, graphical models~\cite{Barber}, or role discovery~\cite{Rossi} that are somewhat similar to our proposed approach have not been considered in our comparison. However, via the above example, the communication theoretic data analytical approach indeed demonstrates its potential, particularly as a way of fusing information, while it seems more difficult to achieve a similar purpose by pure multivariate analysis. Some remaining open problems for communication theoretic data analytics are

\begin{itemize}
\item How to measure the information transfer from each variable $X_i \rightarrow Y$, $i=1,2, \ldots$.
\item How to determine a sufficient amount of information transfer? And thereby, how to determine which variables should be considered in data analytics.
\end{itemize}

A more realistic but complicated scenario involves information fusion and information diversity at the same time, which is a multiuser detection or MIMO problem. Due to space limitations, it is not possible to explore this idea here, in spite of its potential applicability to different problems, such as recommender systems etc. Some open issues include:

\begin{itemize}
\item Joint prediction of $Y_1,Y_2, \ldots ,Y_N$ from $X_1,X_2, \ldots ,X_M$, and associated techniques to effectively solve this problem, such as sub-space approaches etc. 
\item Optimal MUD has NP-hard complexity, leading to suboptimal receivers such as the de-correlating receiver. Comparisons of such structures to multivariate time series analysis, mathematically and numerically, are of interest.
\end{itemize}

\section{Information Coupling}

Thus far, we have intuitively viewed communication theoretic data analytics as being centered on information transfer among different data variables, and then applied receiver techniques to enhance data analytics. This suggests that the amount of information in the data is in fact far less then the amount of (big) data. The methodology in Section III deals with a number of data variables to effectively execute information processing analoguesly to communication systems and multiuser communications. The remaining challenge is to identify low-dimensional information structure from high-dimensional (raw) data, and to better construct intuition about communication theoretic data analytics from information theory. In this section, we aim to apply the recently developed information coupling~\cite{HuangZheng} to achieve this purpose, and also provide information theoretic insights to information-centric data processing.

\subsection{Introduction to Information Coupling}

From the analog between communication networks and data analytics, intuition suggests the need for information-centric processing in addition to data processing for not only optimizing communication systems/networks but also mining important information from (big) data. The conventional studies of information processing are limited to data processing, thereby focusing on representing information as bits, and transmitting, storing, and reconstructing these bits reliably. To see this, let us consider a random variable with $M$ possible values $\{1,2, \ldots , M\}$. If we know its value reliably, then we can describe this knowledge with a single integer, and then further process the data with this known value. On the other hand, if the value is not deterministically acquirable, then we need to describe our knowledge with an $M$-dimensional distribution $P_M (m)$, which requires $M-1$ real numbers to describe. Therefore, the data processing task has to be performed in the space of probability distributions.

When we move towards information-centric processing, the general way to describe information processing relies on the conditional distribution of the message, conditioned on all the observations, at each node of the network, e.g., $P_{Y_n|X_m}$ in Figure~\ref{fig:figure_1}. Conventional information theoretic approaches working on the distribution spaces in communication and data processing are mostly based on coded transmission, in which the desired messages are often quite large, which results in the extremely high dimensionality of the belief vectors. This is in fact one of the main difficulties of shifting the data processing from data centric to information centric. It turns out that this difficulty comes from the fact that the distribution space itself is not a flat vector space, but is a rather complicated manifold. Amari's work~\cite{SH00} on information geometry provides a tool to study this space, but the analysis can be quite involved in many cases. In this section, we propose a framework that allows us to greatly simplify this challenge. In particular, we turn our focus to low rate information contained in the data, which is significant for describing the data. We call such problems information coupling problems~\cite{HuangZheng}\cite{SL08}. 

To formulate this problem mathematically, let us consider a point-to-point communication scenario, where a signal $X$ is transmitted through a channel with the transition probability $W_{Y|X}$, which can be viewed as a $|\cY| \times |\cX|$ matrix, to generate an output $Y$. In the conventional communication systems, we consider encoding a message $U$ into the signal vector $X$, to form a Markov relation $U \rightarrow X \rightarrow Y$. From which, an efficient coding scheme aims to design both the distribution $P_U$ and the conditional distributions $P_{X|U=u}$ to maximize the mutual information $I(U;Y)$, which corresponds to the communication rate. Such optimization problems in general do not have analytical solutions, and require numerical methods such as the Blahut-Arimoto algorithm to find the optimal value. More importantly, when we allow coded transmissions, i.e., to replace $X$ and $Y$ by $n$ independent and identically distributed (i.i.d.) copies of the pair, it is not clear a priori that the optimizing solution would have any structure. Although Shannon provided a separate proof for the point-to-point case that the optimization of the multi-letter problem over $P_{X^n|U}$ should also have an i.i.d. structure, failure to generalize this proof to multi-terminal problems remains the biggest obstacle to solving network capacity and subsequently design algorithms. In contrast, the information coupling deals with the maximization of the same objective function $I(U;Y)$, but with an extra constraint that the information encoded in $X$, measured by $I(U;X )$ is small. With a slight strengthening this constraint can be reduced to the condition that all the conditional distributions $P_{X|U} (\cdot | u)$, for a $u$, are close to the marginal distribution $P_X$. We refer the reader to~\cite{HuangZheng} for the details of this strengthening. With this extra constraint, the linear information coupling problem for the point-to-point channel can be formulated as
\begin{align}\label{eq:LICP}
\max_{U \rightarrow X \rightarrow Y } &\frac{1}{n} I(U;Y),\\ \notag 
\mbox{subject to:} \ \ &\frac{1}{n} I(U;X) \leq \delta, \\  \label{eq:LICP_constraint_1}
\ \ & \frac{1}{n}  \| P_{X|U=u} - P_{X} \|^2 = O(\delta), \ \forall u,
\end{align}
where $\delta$ is assumed to be small.

It turns out that the local constraint~\eqref{eq:LICP_constraint_1} in~\eqref{eq:LICP} that assumes all conditional distributions are close to the marginal distribution, plays the critical role of reducing the manifold structure into a linear vector space. In addition, the optimization problem, regardless of the dimensionality, can always be solved analytically with essentially the same routine. In order to show how the local constraint helps to simplify the problem, we first note that given the conditional distributions $P_{X|U=u}$ are closed to $P_X$ for all $u$ in terms of $\delta$, the mutual information $I(U;X)$ can be approximated up to the first order as
\begin{equation}\label{eq:constraint}
I(U;X) =  \delta \cdot \sum_{u} P_U (u) \cdot \| \psi_u  \|^2 + o(\delta),
\end{equation}
where $\psi_u$ is the perturbation vector with the entries $\psi_u (x) = \left( P_{X|U=u} (x) - P_X (x) \right) / \sqrt{\delta \cdot P_X(x)}$, for all $x$. This local approximation results from the first order Taylor expansion of the Kullback-Leibler (K-L) divergence $D( P_{X|U=u} \| P_X)$ between $P_{X|U=u}$ and $P_X$ with respect to (w.r.t.) $\delta$. In addition, with this approximation technique, we can similarly express the mutual information at the receiver end as
\begin{align}\label{eq:constraint}
I(U;Y) =  \delta \cdot \sum_{u} P_U (u) \cdot \| \hat{\psi}_u  \|^2 + o(\delta),
\end{align}
where $\hat{\psi}_u (y) = \left( P_{Y|U=u} (y) - P_Y (y) \right) / \sqrt{\delta \cdot P_Y(y)}$. Now, note that $U \rightarrow X \rightarrow Y$ forms a Markov relation, therefore both $P_{Y|U=u}$ and $P_Y$, viewed as vectors, are the output vectors of the channel transition matrix $W_{Y|X}$ with the input vectors $P_{X|U=u}$ and $P_X$. This implies that the vector $\hat{\psi}_u$ is the output vector of a linear map $B$ with $\psi_u$ as the input vector, where
\begin{align} \label{eqn:B}
B \triangleq \left[\sqrt{P_Y}^{-1} \right] W_{Y|X} \left[\sqrt{P_X} \right],
\end{align}
and $\left[\sqrt{P_X} \right]$ and $\left[\sqrt{P_Y} \right]$ denote diagonal matrices with diagonal entries $P_X (x)$ and $P_Y (y)$. This linear map $B$ is called the divergence transition matrix (DTM) as it carried the K-L divergence metric from the input distribution space to the output distribution space. 

\begin{figure}[t]
\centering 
\def\svgwidth{1.1\columnwidth}
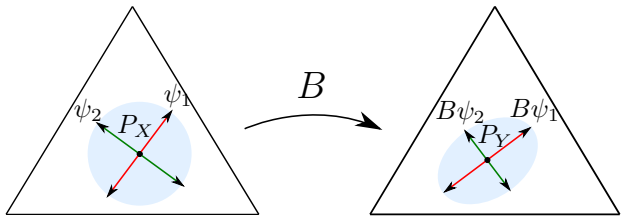
\caption{The divergence transition matrix $B$ serves as a linear map between two spaces, with right and left singular vectors as orthonormal bases. Different input singular vectors have different output lengths at the output space.}
\label{fig:P2P_Coupling_1}
\end{figure}

We shall point out here that with this local approximation technique, both the input and output probability distribution spaces are linearized as Euclidean spaces by the tangent planes around the input and the output distributions $P_X$ and $P_Y$. Hence, we can define the coordinate system in both distribution spaces, such as the inner product and orthonormal basis, as in the conventional Euclidean spaces. Under such a coordinate system, the mutual information $I(U;X)$ becomes the Euclidean metric of the perturbation vector $\psi_u$ averaged over different values of $u$. Similarly, the mutual information $I(U;Y)$ can also be viewed as the Euclidean metric of the perturbation vector $B \cdot \psi_u$ at the output space. Hence, the optimization problem of maximizing the mutual information $I ( U ; Y )$ is turned into the following linear algebra problem:
\begin{align}\label{eq:bbb200}
\max. \ &\sum_{u} P_U(u) \cdot \left\| B \cdot \K_u \right\|^2 \\ \notag
\mbox{subject to:} \ &\sum_{u} P_U(u) \cdot \| \K_u \|^2 = 1.
\end{align}
In particular, $U$ can without loss of the optimality be designed as a uniform binary random variable, and the goal of~\eqref{eq:bbb200} is to find the input perturbation vector $\psi_u$ that provides the largest output image $B \cdot \psi_u$ through the linear map $B$. The solution of this problem then relies on the singular value decomposition of $B$, and the optimal $\psi_u$ corresponds to the singular vector of $B$ w.r.t. the largest singular value. Figure~\ref{fig:P2P_Coupling_1} illustrates the geometric intuition of linearized distributions spaces.

More importantly, this information coupling framework with the locality constraint allows us to deal with the multi-letter problems in information theory in a systematic manner. To see this, consider the multi-letter version of the problem~\eqref{eq:LICP} 
\begin{align}\label{eq:LICP_multi}
\max_{U \rightarrow X^n \rightarrow Y^n } &\frac{1}{n} I(U;Y^n),\\ \notag 
\mbox{subject to:} \ \ &\frac{1}{n} I(U;X^n) \leq \delta, \\ \notag
\ \ & \frac{1}{n}  \| P_{X^n|U=u} - P_{X^n} \|^2 = O(\delta), \ \forall u,
\end{align}
in which the message is encoded in an $n$-dimensional signal vector $X^n$ and the optimization is over the distribution space of $P_{X^n | U=u}$. By applying the same local approximation technique, we can again linearize both input and output spaces into Euclidean spaces, and the linear map between these two spaces turns out to be the tensor product $B^n$. Due to the fact that the singular vectors of $B^n$ are tensor products of the singular vectors of $B$, we know that the optimal perturbation vector in the multi-letter case has the tensor product form, and the optimal conditional distribution $P_{X^n | U=u}$ has an i.i.d. structure~\cite{HuangZheng}. More interestingly, this approach of dealing with multi-letter information theory problems can be easily carried to multi-terminal problems, in which all the information theory problems are simply reduced to the corresponding linear algebra problems. In particular, the i.i.d. structure of $P_{X^n | U=u}$ for the point-to-point case was also observed by Shannon with an auxiliary random variable approach; however, the generalization of the auxiliary random variable approach to multi-terminal problems, e.g., the general broadcast channel, turns out to be difficult open problems. In a nutshell, the information coupling and the local constraint help us to reduce the manifold structure into a linear vector space, where the optimization problem, regardless of the dimensionality can always be solved analytically with essentially the same routine.

Furthermore, the information coupling formulation not only simplifies the analysis, but also suggests a new way of communication over data networks or information transfer over networks of data variables. Instead of trying to aggregate all the informations available at a node, pack them into data packets, and send them through the outgoing links, the information coupling methodology seeks to transmit a small piece of information at a time, riding on the existing data traffic~\cite{HuangSuhZheng}. The network design of data variables thus focuses on the propagation of a single piece of message, from the source data variable to all destination data variables. Each node in the network only alters a small fraction of the transmitted symbols, according to the decoded part of this message. The analytical simplicity of the information coupling allows such transmissions to be efficient, even in the presence of general broadcasting and interference. Furthermore, information coupling can be employed to obtain useful information from network operation, as a complementary function for (wireless) network tomography. Consequently, we can analyze the covariance matrix of received signals at the fusion center in a sensor network to form communities like social networks such that energy efficient transmission and device management can be achieved.

\subsection{ Implementation of Information Coupling}

Using the information theoretic setup via information coupling, we shall demonstrate how to deal with practical data analytics. To infer useful results from big data, we shall be able to acquire important knowledge in general social network modeling of big data such as Figure~\ref{fig:figure_1}. In particular, we consider $X_1 , \ldots , X_M$ as information transmitters and $Y_1 , \ldots , Y_N$ as the information receivers. The probabilistic relationship between $X$'s and $Y$'s represents the communication channel, which copes with the effects of imperfect sampling, noisy observation, or interference from unknown variables or outliers. In the following, we are going to demonstrate the potential of extracting critical low dimensional information from (big) data through the innovative information coupling approach.

To demonstrate the idea, suppose that there is a hidden source sequence $\ux = \{ x_1, x_2, \ldots , x_n \}$, i.i.d. generated according to some distribution $P_X$. Instead of observing the hidden source directly, we are only allowed to observe a sequence $\uy = \{ y_1, y_2, \ldots , y_n \}$, which can be statistically viewed as the noisy outputs of the source sequence through a discrete memoryless channel $W_{Y|X}$. Traditionally, if we want to infer the hidden source from the noisy observation $\uy$, we would resort to a low-dimensional sufficient statistic of $\uy$ that has all the information one can tell about $\ux$. However, in many cases, such a sufficient statistic might be computationally difficult to obtain due to the high dimensional structures of $\ux$ and $\uy$, which turns out to be the common obstacle in dealing with big data. In contrast to seeking a useful sufficient statistic, we would like to rather turn our focus to consider the statistic from $\uy$ that is \emph{efficient} to describe a certain feature of $\ux$. 

In particular, there are many different ways to define the efficiency of information extraction from data. From the information theoretic point of view, we would like to employ the mutual information as the measurement of the information efficiency about the data. Rigorously, we want to acquire a binary feature $U$ in $\ux$ from the observed data $\uy$, such that the efficiency, measured by $I(U; Y^n)$, can be maximized. In order to find such a feature, we shall formulate an optimization problem that has the same form as the linear information coupling problem~\eqref{eq:LICP_multi}, and the optimal solution of~\eqref{eq:LICP_multi} characterizes which feature of $\ux$ can be the most efficiently extracted from the noisy observation $\uy$ in terms of the mutual information metric. Therefore, from~\cite{HuangZheng}, we can explicitly express the optimal solution $P_{X^n|U}$ of~\eqref{eq:LICP_multi} as the tensor product of the distribution $P_{X|U} (x) = P_{X} (x) \pm \sqrt{\delta P_X (x)} \cdot \psi_X (x) $, where $\psi_X$ is the singular vector of the DTM with the largest singular value.

\begin{figure}[t]
\centering 
\def\svgwidth{\columnwidth}
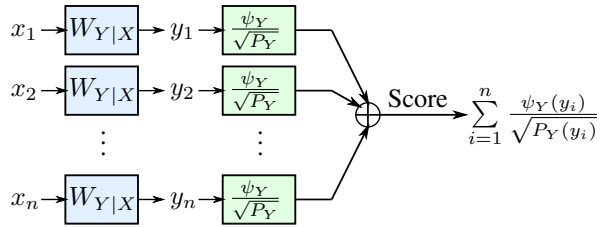
\caption{The score function for the noisy observations.}
\label{fig:p2p}
\end{figure}

Then, we want to estimate this piece of information $U$ from the noisy observation $\uy$. For this purpose, we apply the maximum likelihood principle, and the log-likelihood function can be written as
\begin{align*} 
l_i(\uy) = \log \left( \frac{P_{\uY|U=i}(\uy)}{P_{\uY}(\uy)} \right), \quad i = 0,1,
\end{align*}
where $P_{\uY}$ and $P_{\uY|U}$ are the output distributions of the channel $W_{Y|X}$ with input distributions $P_{\uX}$ and $P_{\uX|U}$. Then, the decision rule depends on the sign of $l_0(\uy) - l_1(\uy)$: when it is positive, we estimate $\hat{U} = 0$, otherwise, $\hat{U} = 1$. Now, noting that both $P_{\uY|U}$ and $P_{\uY}$ are product distributions, we can further simplify $l_0(\uy)$ as
\begin{align*} 
l_0(\uy) 
&= \sum_{i = 1}^n \log \left( \frac{P_{Y|U=i}(y_i)}{P_{Y}(y_i)} \right) \\
&= \sum_{i = 1}^n \log \left( 1 + \sqrt{\delta} \cdot \frac{\psi_{Y}(y_i)}{\sqrt{P_Y(y_i)}} \right) \\
&\simeq \sqrt{\delta} \cdot \sum_{i = 1}^n \frac{\psi_{Y}(y_i)}{\sqrt{P_Y(y_i)}},
\end{align*}
where in the last equation, we ignore all the higher order terms of $\delta$. 
We call $\frac{\psi_{Y}}{\sqrt{P_Y}}$ the \emph{score function} $\frac{\psi_{Y}}{\sqrt{P_Y}} : {\cal Y} \mapsto \mathbb{R}$, in which the empirical sum of this function over the data $y_1 , \ldots , y_n$ is the sufficient statistic of a specific piece of information in $\ux$ that can be the most efficiently estimated from $\uy$. Figure~\ref{fig:p2p} illustrates the score function in this point-to-point setup. 
The score function derived from the information coupling approach provides the maximal likelihood statistics of the most efficiently inferable information from the data, and we call the score function the \emph{efficient statistic} of the data. The efficient statistic of the data can be deemed as a low dimensional label corresponding to the most significant information of the data that can be employed in further data processing tasks. In the next subsection, we shall demonstrate how to apply the efficient statistic to practical machine learning problems and its performance through an image recognition example.

Finally, we would like to emphasize that the efficient statistic can be useful in many machine learning scenarios, such as image processing, network data mining and clustering. Consider the social network modeling of big data as Figure~\ref{fig:figure_1} with very large number of nodes in the network. In this case, acquiring a meaningful sufficient statistic for the data is usually an intractable task due to the complicated network structure. Moreover, even if it is possible to specify the sufficient statistic, the computational complexity can still be extremely high due to the high dimensional structure of the data. On the other hand, the efficient statistic obtained from the information coupling provides the information that, while low dimensional, keeps the most significant information about the original data. This is precisely the main objective of the dimension reduction or feature extraction studied in machine learning subjects. Equalization in Section III may be considered as an intuitive implementation of information coupling in big data. In addition, in order to acquire the efficient statistic from the data, we simply need to solve the score function, i.e., the optimal singular vector, which can be computationally efficient. Therefore, we could see that information coupling potentially provides a new framework for efficiently processing and analyzing big networked data.

\subsection{Application to Dimension Reduction in Pattern Recognition} \label{sec:app}

\begin{figure}
\centering 
\subfigure[]{
\includegraphics[width=\columnwidth]{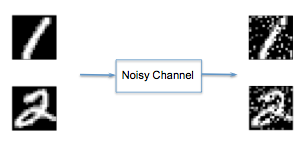}
\label{fig:Handwriting}
}
\subfigure[]{
\def\svgwidth{\columnwidth}
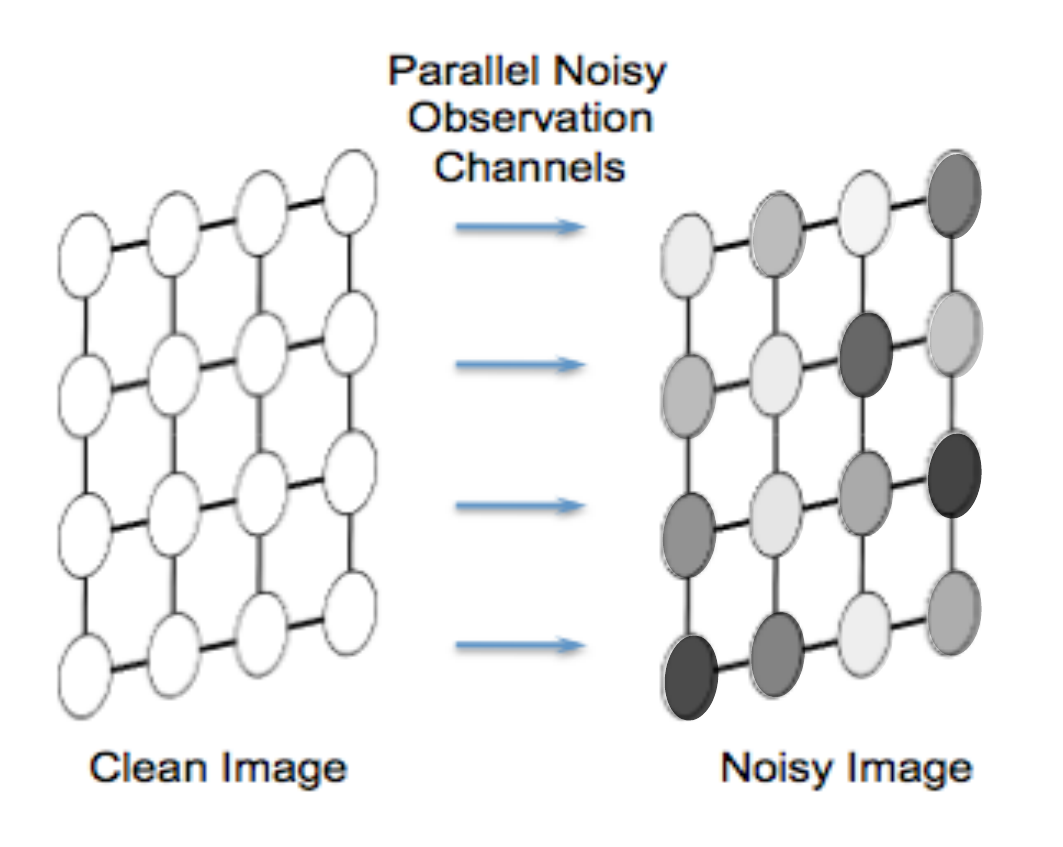
\label{fig:Ising}
}
\caption{(a) Handwriting Recognition between the number ``1" and ``2" via the noisy images. (b) Ising Model of Noisy Images. The pixels of the clean image can be viewed as random variables $X_i$. After passing through the channel, the pixels are corrupted by the noise to different levels, and the collection of noisy pixels are the random variables $Y_i$.}
\end{figure}

Let us illustrate how the efficient statistic can be applied to practical data processing. For demonstration purposes, we aim to address the image recognition task of handwriting numbers ``1" and ``2" through noisy images, as illustrated in Figure~\ref{fig:Handwriting}. We consider these 2-D images as from an Ising model as shown in Figure~\ref{fig:Ising}. Each clean pixel in Figure~\ref{fig:Ising} is passed through one of the parallel independent noisy observation channels, to get a noisy image. In abstract, we can think of the pixels of the clean image as a collection of random variables $X_1, X_2 , \ldots , X_N$. Then, passing through noisy observation channels with a transition kernel $P_{Y^N | X^N}$, the pixels of the noisy image is a collection of random variables $Y_1, Y_2 , \ldots , Y_N$. Now, to apply the efficient statistic to acquire the most significant feature, we shall in principle go through the following procedures:
\begin{itemize}
\item [(1)] To determine the divergence transition matrix $B$, we shall determine the distributions $P_{X^N}$ and $P_{Y^N}$ as well as the transition kernel $P_{Y^N | X^N}$. The distributions $P_{X^N}$ and $P_{Y^N}$ can be learned from the empirical distribution of the images viewed as $N$-dimensional vectors. In addition, we design the transition kernel $P_{Y^N | X^N}$ in this image recognition example.
\item [(2)] Solve the singular value decomposition of $B$ and determine the optimal left singular vector $\psi$ of $B$. Note that $\psi$ has the dimensionality $| \cY |^N$.
\item [(3)] The efficient statistic is then specified by the score function $\psi(y^N)/ P_{Y^N}(y^N)$ of the data $y^N = \{ y_1, \ldots, y_N \}$, which can be obtained by the $y^N$-th entry of the vector $\psi_Y$, divided by $P_{Y^N}(y^N)$.
\end{itemize}

Now, let us demonstrate the application of this procedure to a practical image recognition problem. Here, we employ the MNIST Database~\cite{MNIST} of handwritten digits as the test images, where each image is of size $19\times19$ pixels and each pixel value is scaled to have one of four values as in Figure~\ref{fig:Handwriting}. In particular, as shown in Figure~\ref{fig:Handwriting}, we have a mixture of images of handwritten digits $1$ and $2$, and we assume that we can observe the noisy version of these images. Our goal is to separate these noisy images with respect to different digits by computing the score of these images with our algorithm and ordering them.

\begin{figure}
\centering 
\def\svgwidth{\columnwidth}
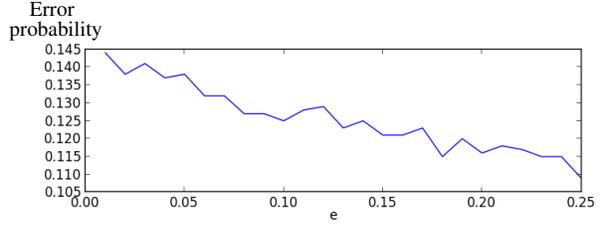
\caption{Experimental results for the separation efficiency with respect to different values of $e$.}
\label{fig:eff} 
\end{figure}

To this end, we view the (clean) images as generated from the Ising model, in which each pixel corresponds to the nodes $X_i$ in the Ising model. Then, we pass each pixel in the images independently through a discrete memoryless channel with transition matrix  
\begin{align} \label{eq:channel_matrix}
\left[\begin{matrix}
1-2e & 2e & e & \frac{e}{2} \\
e & 1-3e & 2e & \frac{e}{4} \\
e & 0 & 1-4e & \frac{e}{4} \\
0 & e & e & 1-e
\end{matrix}\right].
\end{align}
Here, the transition matrix is chosen merely to facilitate our simulation, and $e$ is the parameter measuring the noise level of the channel. After passing clean pixels through the channel, we observe the noisy version of these images, where each noisy pixel corresponds to $Y_i$ in this setup.  Clearly, the empirical joint input and output distributions can be obtained by the statistics of the images. Then, we can apply our algorithm to compute the score for each noisy image, and then order these scores to separate images with respect to different digits.

To measure the performance of our algorithm, we classify a batch of $2N$ images with $N$ of $1$'s and $N$ of $2$'s. After ordering the scores, an ideal classifier should have the $N$ lowest scored images belonging to one digit and the $N$ largest scores belonging to the other digit. To compare with the ideal classifier, we define the separation error probability as the proportion of the pictures that is wrongly classified, i.e. 
\begin{align}
\text{Error probability} = \frac{\text{$\#$ of wrongly classified pictures}}{2N}
\end{align}
The classifier is more efficient when the separation factor is closer to $0$. For different values of $e$, our algorithm has the performance as in Figure~\ref{fig:eff}. From this simulation result, we can see that our algorithm is quite efficient in separating images with respect to different digits. This result tells that the efficient statistic is in fact a very informative way to describe stochastic observations.

\begin{remark}
It might be curious at first glance that in Figure~\ref{fig:eff}, the error probability does not decay as the noise level $e$ grows. In fact, this phenomenon can be explained as follows. Note that the score function defined in this section not only depends on the data vectors $X^N$, but also on the designed channel transition matrix~\eqref{eq:channel_matrix}. Therefore, different channel transition matrices may provide different score function on the noisy data vectors $Y^N$. We shall notice that the score function is designed to extract the feature that can be communicated the best through the channel, but not necessary the best feature to separate the two sets of images. Thus, the performance of the image recognition may not be improved with a less noisy channel. On the other hand, we should understand our result as that, with a rather arbitrarily designed channel transition matrix~\eqref{eq:channel_matrix}, we have obtained a rather nice performance of error probabilities, which does not require any extra learning other than the empirical distributions of the data, i.e., completely unsupervised. Thus, our result demonstrates a new potential of applying communication and information theory to machine learning problems.
\end{remark}

\begin{remark}
Dimension reduction is one of the central topics in statistics, machine learning, pattern recognition, and data mining, and has been studied intensively. Celebrated techniques addressing this subject including principal component analysis (PCA)~\cite{Jolliffe}, $K$-means clustering~\cite{DingHe}, independent component analysis (ICA)~\cite{AJ01}, and regression analysis~\cite{Freedman}, where many efficient algorithms have been developed to implement these approaches~\cite{Kalman}-\cite{Koller}. In particular, these approaches mainly focus on dealing with the space of the data, rather than addressing the information flow embedded in the data. On the other hand, recent studies have suggested the trend of information-centric data processing~\cite{3}, thus advocating the research direction of analyzing the underlying information flow of networked data. The information coupling approach can be considered as a technique that aims to provide a framework to reach this goal from the information theoretic perspective. From the discussions in this section, we can see that information coupling studies the data analysis problems from the angle of distribution space but not simply the data space, Thus, information coupling potentially provides a fresh view of how information can be exchanged between different terminals in implementing the data processing tasks, which not only helps to more deeply understand the existing approaches, but also opens a new door to develop new technologies. 
\end{remark}

\begin{remark}
While this simple image recognition example illustrates the feasibility of introducing information coupling to data analysis problems, there are critical challenges for future research:
\begin{itemize}
\item How to develop efficient iterative algorithms that exploit the structure of the graphical models to compute the singular vectors and evaluate the scores.
\item In the case where some training data are available, how the information coupling approach can be adjusted to cooperate with the side information.
\item Except for the most informative bit, how can we extract the second and third bits from the data, and how these bits can be applied to deal with practical data analysis tasks.
 \end{itemize}
\end{remark}

\section{Conclusions}

Statistical analysis on big data has usually been treated as an exercise in statistical data processing. With the help of statistical communication theory, we have introduced a new methodology to enable information-centric processing (or statistical information processing) for big data. Hopefully, this opens new insights into both big data analytics and statistical communication theory. 

Although we have demonstrated initial feasibility of this methodology, there are further critically associated challenges ahead, namely

\begin{itemize}
\item How to identify appropriate or enough variables to influence one variable (or a set of variables).
\item How to detect outliers~\cite{Tajer}.
\item How to generalize big data analytics using large communication network analysis beyond multiuser communications.
\item How to interpret and adopt traditional machine learning approaches and data processing technologies, such as (un)supervised learning, feature selection, blind source separation, via the techniques developed in network communication theories.
\end{itemize}

\appendices
\section{Equalizer Implementation for Communication Theoretic Data Analytics}

As in Fig. 4, by proper selection of adaptive algorithm and step size, the output of the equalizer after training period gives the inference
\begin{align} \label{A1}
\hat{y} [n] = \sum_{l = 0}^L w_l x[n-l].
\end{align}
Based on the minimum MSE criterion, the purpose of training data is to obtain
\begin{align} \label{A2}
\mathop{\arg\min}_w \left\{ \left( y[n] - \sum_{l=0}^L w_l x[n-l] \right)^2  \right\},
\end{align}
and
\begin{align*}
\mathop{\arg\min}_L \left\{ \left( y[n] - \sum_{l=0}^L w_l x[n-l] \right)^2  \right\}.
\end{align*}
The first equation is to obtain the vector of weighting coefficients, and the second equation is to identify the most appropriate observation depth, $L$. Once we identify $L$, we keep it and therefore the equalizer structure to infer data. We may keep the same set of coefficients or update online. Please note we may also obtain a predictor as follows:
\begin{align} \label{A3}
\hat{y} [n+1] = \sum_{l=0}^L w_l x[n-l],
\end{align}
where we cannot go into further detail due to the length constraint on this paper.

\begin{figure} 
\centering 
\includegraphics[width=\columnwidth]{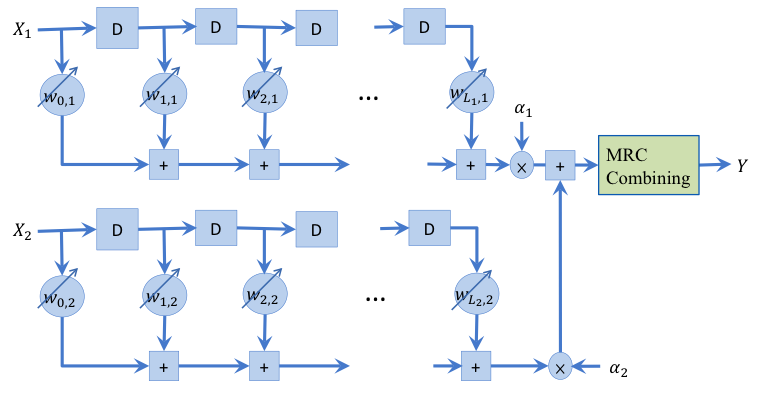} 
\caption{Maximal-Ratio Combining of Two Equalized Data Variables (Time Series).}
\label{fig:figure_A}
\end{figure}

When we have two (or more) data variables to infer another data variable, say using $X_1$ and $X_2$ to infer $Y$, we have to use information fusion as in Figure~\ref{fig:figure_A}. Again, we adopt the minimum MSE criterion, to yield
\begin{align} \label{A4}
\min_{\alpha_m[n]} E \left\{ \left( Y - \sum_{m=1}^M \alpha_m[n] \hat{y}_m [n] \right)^2  \right\},
\end{align}
where
\begin{align*}
\hat{y}_m [n] = \sum_{l=0}^{L_m} w_{l,m} x_m [n-l] .
\end{align*}
Necessary conditions for this minimization gives the solution for $\alpha_m[n]$. The consequent estimator is therefore
\begin{align} \label{A5}
\hat{y} [n] = \sum_{m=1}^M \alpha_m[n] \hat{y}_m [n] = \sum_{m=1}^M \alpha_m[n] \sum_{l=0}^{L_m} w_{l,m} x_m [n-l]
\end{align}
which is defined as the maximal ratio combining of equalized multivariate regression of different optimal observation lengths $L_m$, $m=1, \ldots ,M$. This design realizes the idea of maximizing information flow between data variables or time series.  For ease of implementation, we may set $\alpha_m[n] = \alpha_m$, or we may adopt selective combining and equal-gain combining. 

\begin{remark} 
A conjecture to explain why we intend to equalize data of a certain length $L_m$, instead of the entire data set, is that earlier components in the time series may introduce very noisy information, like interference or noise in multiuser communication systems or simply weakly correlated information after a large time separation. Such lengths $L_m$, $m=1,. \ldots ,M$, for data variables $X_1, \ldots ,X_M$, represent the span/range of useful data for inference. Of course, based on the MSE, we may further select useful data variables among $X_1, \ldots ,X_M$. Similar concepts are not rare in machine learning, for example, to identify support vectors in support vector machines (SVMs). What we are doing here is more effective implementation by properly selecting data variables, range of observations, and finally weighting coefficients in each equalizer, for multivariate-regression leveraging the optimization of information transfer between relational data variables. 
\end{remark}

\section*{Acknowledgment}

The authors thank Jia-Pei Lu of National Taiwan University, and Fabi\'an Kozynski of Massachusetts Institute of Technology for their programming of the numerical examples.

\vspace{-15 mm}
\begin{IEEEbiography}
[{\includegraphics[width=1in,height=1.25in,clip,keepaspectratio]{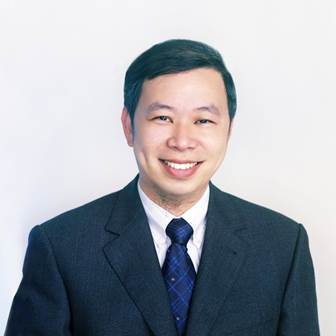}}]{Kwang-Cheng Chen}
(M'89-SM'94-F'07) Kwang-Cheng Chen (MÕ89-SMÕ94-FÕ07) received the B.S. from the National Taiwan University in 1983, and the M.S. and Ph.D from the University of Maryland, College Park, United States, in 1987 and 1989, all in electrical engineering. From 1987 to 1998, Dr. Chen worked with SSE, COMSAT, IBM Thomas J. Watson Research Center, and National Tsing Hua University, in mobile communications and networks. Since 1998, Dr. Chen has been with National Taiwan University, Taipei, Taiwan, ROC, and is the Distinguished Professor and Associate Dean for academic affairs in the College of Electrical Engineering and Computer Science, National Taiwan University. He has been actively involving in the organization of various IEEE conferences as General/TPC chair/co-chair, and has served in editorships with a few IEEE journals. Dr. Chen also actively participates in and has contributed essential technology to various IEEE 802, Bluetooth, and LTE and LTE-A wireless standards. Dr. Chen is an IEEE Fellow and has received a number of awards such as the 2011 IEEE COMSOC WTC Recognition Award, 2014 IEEE Jack Neubauer Memorial Award and 2014 IEEE COMSOC AP Outstanding Paper Award. Dr. His recent research interests include wireless communications, network science, and data analytics.
\end{IEEEbiography}
\begin{IEEEbiography}
[{\includegraphics[width=1in,height=1.25in,clip,keepaspectratio]{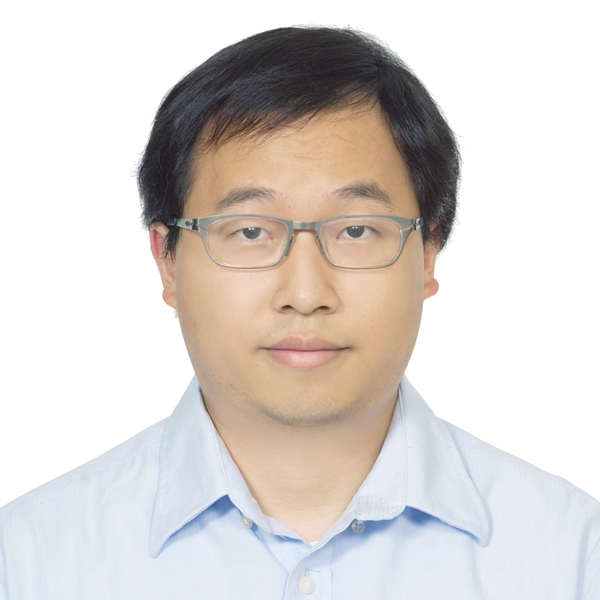}}]{Shao-Lun Huang}
received the B.S. degree with honor in 2008 from the Department of Electronic Engineering, National Taiwan University, Taiwan, and the M.S. and Ph.D. degree in 2010 and 2013 from the Department of Electronic Engineering and Computer Sciences, Massachusetts Institute of Technology. Since 2013, he has been working jointly in the Department of Electrical Engineering at the National Taiwan University and the Department of Electrical Engineering and Computer Science at the Massachusetts Institute of Technology, where he is currently a postdoctoral researcher. His research interests include communication theory, information theory, machine learning and social networks.
\end{IEEEbiography}
\vspace{-41 mm}
\begin{IEEEbiography}
[{\includegraphics[width=1in,height=1.25in,clip,keepaspectratio]{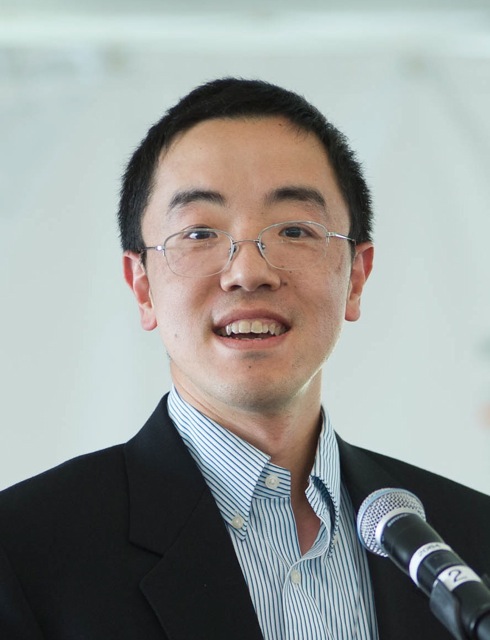}}]{Lizhong Zheng}
received the B.S and M.S. degrees, in 1994 and 1997 respectively, from the Department of Electronic Engineering, Tsinghua University, China, , and the Ph.D. degree, in 2002, from the Department of Electrical Engineering and Computer Sciences, University of California, Berkeley. Since 2002, he has been working in the Department of Electrical Engineering and Computer Sciences, where he is currently a professor of Electrical Engineering and Computer Sciences. His research interests include information theory, wireless communications, and statistical inference. He received the IEEE Information Theory Society Paper Award in 2003, and NSF CAREER award in 2004, and the AFOSR Young Investigator Award in 2007. 
\end{IEEEbiography}
\vspace{-42 mm}
\begin{IEEEbiography}
[{\includegraphics[width=1in,height=1.25in,clip,keepaspectratio]{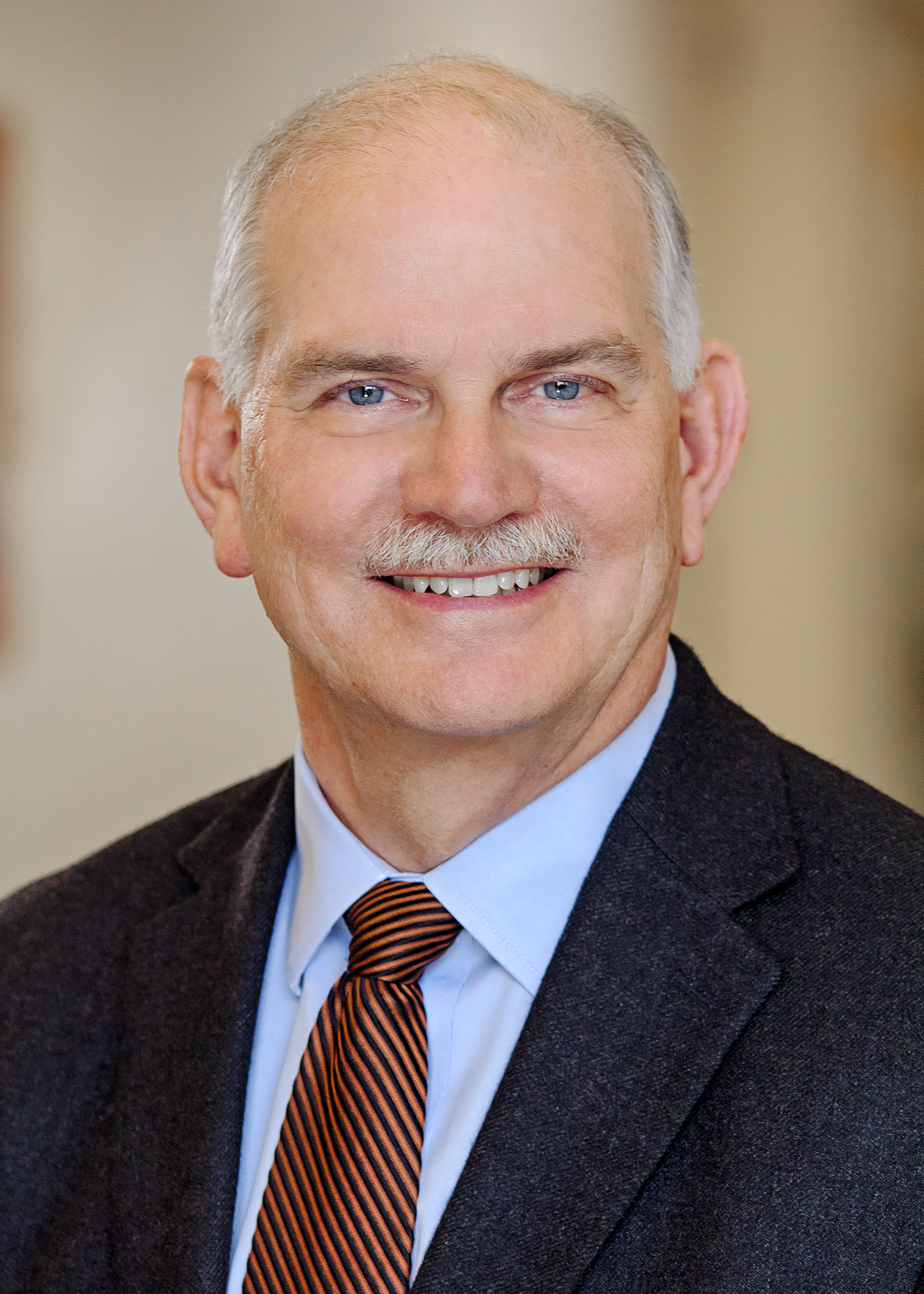}}]{H. Vincent Poor}
(S'72, M'77, SM'82, F'87) received the Ph.D. degree in EECS from Princeton University in 1977. From 1977 until 1990, he was on the faculty of the University of Illinois at Urbana-Champaign. Since 1990 he has been on the faculty at Princeton, where he is the Michael Henry Strater University Professor of Electrical Engineering and Dean of the School of Engineering and Applied Science. Dr. Poor's research interests are in the areas of information theory, statistical signal processing and stochastic analysis, and their applications in wireless networks and related fields including social networks and smart grid. Among his publications in these areas are the recent books \emph{Principles of Cognitive Radio} (Cambridge University Press, 2013) and \emph{Mechanisms and Games for Dynamic Spectrum Allocation} (Cambridge University Press, 2014).

Dr. Poor is a member of the National Academy of Engineering, the National Academy of Sciences, and is a foreign member of Academia Europaea and the Royal Society. He is also a fellow of the American Academy of Arts and Sciences, the Royal Academy of Engineering (U. K), and the Royal Society of Edinburgh. He received the Marconi and Armstrong Awards of the IEEE Communications Society in 2007 and 2009, respectively. Recent recognition of his work includes the 2014 URSI Booker Gold Medal, and honorary doctorates from Aalborg University, Aalto University, the Hong Kong University of Science and Technology, and the University of Edinburgh.
\end{IEEEbiography}

\begin{thebibliography}{1}


\bibitem{1}
X.Wu, X. Zhu, G.Q. Wu, W. Ding, ``Data Mining with Big Data," {\em IEEE Trans. On Knowledge and Data Engineering}, vol. 26, no. 1, pp. 97-107, Jan. 2014.

\bibitem{2}
M.J. Wainwright, M.I. Jordan, ``Graphical Models, Exponential Families, and Variational Inference," {\em Foundations and Trends in Machine Learning}, vol. 1, no. 1-2, pp. 1-305, January 2008.


\bibitem{16}
J. Han, M. Kamber, {\em Data Mining: Concepts and Techniques}, 2nd edition, 2006.

\bibitem{17}
U. Fayyad, G. Piatetsky-Shapiro, R. Smyth, ``From Data Mining to Knowledge Discovery in Databases," {\em AI Magazine}, vol. 17, no. 3, pp. 37-54, 1996.


\bibitem{Jensen}
F. V. Jensen and T. D. Nielsen, {\em Bayesian Networks and Decision Graphs, 2nd ed.}, Springer, 2007. 

\bibitem{Koller}
D. Koller and N. Friedman, {\em Probabilistic Graphical Models: Principles and Techniques.}, The MIT Press, 2009. 


\bibitem{18}
T. Hastie, R. Tibshirani, J. Friedman, {\em The Elements of Statistical Learning}, 2nd edition, Springer, 2009. 

\bibitem{19}
R. Silva, R. Scheines, C. Glymour, P. Spirtes, ``Learning the Structure of Linear Latent Variable Models," {\em J. of Machine Learning Research}, vol. 7, pp. 191-246, 2006. 

\bibitem{20}
J.B. Tenenbaum, V. de Silva, J.C. Langford, ``A Global Geometric Framework for Nonlinear Dimensionality Reduction," {\em Science}, vol. 290, pp. 2319-2323, December 22, 2000.

\bibitem{21}
S.T. Roweis, L.K. Saul, ``Nonlinear Dimensionality Reduction by Locally Linear Embedding," {\em Science}, vol. 290, pp. 2323-2326, December 22, 2000.

\bibitem{22}
S. Ji, Y. Xue, L. Carin, ÒBayesian Compressive Sensing," {\em IEEE Trans. On Signal Processing}, vol. 56, no. 6, pp. 2346-2356, June 2008.

\bibitem{23}
T.G. Kolda, B.W. Bader, ``Tensor Decomposition and Applications," {\em SIAM Review}, vol. 51, no. 3, pp. 455-500, 2009.

\bibitem{24}
S.C. Lin, K.C. Chen, ``Improving Spectrum Efficiency via In-Network Computations in Cognitive Radio Sensor Networks," {\em IEEE Trans. on Wireless Communications}, vol. 13, no. 3, pp. 1222-1234, March 2014.

\bibitem{25}
X.-W. Chen, A. Lin, ``Big Data Deep Learning: Challenges and Perspectives," {\em IEEE Access}, vol. 2, pp. 514-525, 2014.

\bibitem{26}
W. Tan, M.B. Blake, I. Saleh, S. Dustdar, ``Social-Network-Sourced Big Data Analytics," {\em IEEE Internet Computing}, pp. 62-69, Sep/Oct 2013. 


\bibitem{3}
K.C. Chen, M. Chiang, H.V. Poor, ``From Technological Networks to Social Networks," {\em IEEE Journal on Selected Areas in Communications}, vol. 31, no. 9, pp. 548-572, September Supplement 2013.


\bibitem{10}
D. Estrin, ``Small Data Where n = me," {\em Communication of the ACM}, vol. 57, no. 4, pp. 32-34, April 2014. 

\bibitem{4}
S. Verd\'u, {\em Multiuser Detection}, Cambridge University Press, 1998. 



\bibitem{Koivisto}
M. Koivisto, K. Sood, ``Exact Bayesian Structure Discovery in Bayesian Networks," {\em J. Machine Learning Research}, vol. 5, pp. 549-573, May 2004.

\bibitem{28}
Z. Wang, L. Chan, ``Learning Causal Relations in Multivariate Time Series Data," {\em ACM Trans. On Intelligent Systems and Technologies}, vol. 3, no. 4, Article 76, September 2012.




\bibitem{5}
G. Stuber, J. Barry, S McLaughlin, G. Li, M. Ingram, T. Pratt, ``Broadband MIMO-OFDM Wireless Communications," {\em IEEE Journal on Selected Areas in Communications}, vol. 92, no. 2, pp. 271-294, Feb. 2004.

\bibitem{Sandryhalia}
A. Sandryhalia, J.M.F. Moura, ``Big Data Analysis with Signal Processing on Graphs," {\em IEEE Signal Processing Magazine}, vol. 31, no. 5, pp. 80-90, September 2014.

\bibitem{6}
R.S. Blum, S.A. Kassam, H.V. Poor, ``Distributed Detection with Multiple Sensors: Part II," {\em Proceeding of the IEEE}, vol. 85, no. 1, pp. 64-79, Jan. 1997.








\bibitem{11}
L. Tong, S. Perreau, ``Multichannel Blind Identification: From Subspace to Maximum Likelihood Methods," {\em Proceeding of the IEEE}, vol. 86, no. 10, pp. 1951-1968, October 1998. 

\bibitem{Yu}
C. Yu, L. Xie, Y.C. Soh, ``Blind Channel and Source Estimation in Networked Systems," {\em IEEE Trans. on Signal Processing}, vol. 62, no. 17, pp. 4611-4626, Sep. 2014


\bibitem{29}
S.M. Kay, {\em Fundamentals of Statistical Signal Processing: Estimation Theory}, Prentice-Hall, 1993.

\bibitem{30}
A.D. Rencher, W.F. Christensen, {\em Methods of Multivariate Analysis}, John Wiley \& Sons, 2012.

\bibitem{31}
L. Lacasa, B. Luque, F. Baillesteros, J. Luque, J.A. Nuno, ``From Time Series to Complex Networks: The Visible Graph," {\em Proceeding National Academy of Science}, vol. 105, no. 13, pp. 4932-4975, April 1 2008.

\bibitem{32}
A.S.L.O. Campanharo, et al., ``Duality Between Time Series and Networks," {\em PLoS ONE}, vol. 6, no. 8, e23378, August 2011. 

\bibitem{33}
M. Last, Y. Klein, A. Kandel, ``Knowledge Discovery in Time Series Databases," {\em IEEE Trans. On Systems, Man, Cybernetics -- Part B: Cybernetics}, vol. 31, no. 1, pp. 160-169, Feb. 2001.

\bibitem{34}
I.S. Dhillon, S. Mallela, D.S. Modha, ``Information-Theoretical Co-clustering," {\em Proceedings of ACM SIGKDD}, 2003.


\bibitem{Esling}
P. Esling, C. Agon, ``Time-series Data Mining," {\em ACM Computing Survey}, vol. 45, no. 1, Article 12, November 2012.


\bibitem{Harvey}
A.C. Harvey, R.G. Pierse, ``Estimating Missing Observations in Economic Time Series," {\em J. American Statistical Association}, vol. 79, no. 385, pp. 125-131, Mar. 1984

\bibitem{Barber}
D. Barber, A.T. Cemgil, ``Graphical Models in Time Series," {\em IEEE Signal Processing Mag.}, Nov. 2010.

\bibitem{Rossi}
R.A. Rossi, N.K. Ahmed, ``Role Discovery in Networks," to appear in the {\em IEEE Tr. on Knowledge and Data Engnieering}.



\bibitem{SH00}
S.-I. Amari, H. Nagaoka,
\newblock {\em Methods of Information Geometry}, Oxford University Press, 2000.

\bibitem{HuangZheng}
S.-L. Huang, L.~Zheng, ``Linear Information Coupling Problems,'' {\em Proceedings of the IEEE International Symposium on Information Theory}, July 2012.  (Full paper submitted to the {\em IEEE Trans. On Information Theory} 2014.)

\bibitem{SL08}
S. Borade, L. Zheng,
\newblock  ``Euclidean Information Theory,"
\newblock {\em Proceedings of the IEEE International Zurich Seminars on Communications}, March, 2008.



\bibitem{HuangSuhZheng}
S.-L. Huang, C.~Suh, L.~Zheng, ``Euclidean Information Theory of Networks," {\em Proceedings of the IEEE International Symposium on Information Theory}, July 2013.





\bibitem{MNIST}
MNIST Handwritten Digit Database,~from~\url{http://yann.lecun.com/exdb/mnist/}

\bibitem{Jolliffe}
I. T. Jolliffe,
\newblock {\em Principal Component Analysis,} Springer Verlag, 1986.

\bibitem{DingHe}
C. Ding, X. He, ``K-means Clustering via Principal Component Analysis," {\em In Proceedings of the Twenty-First International Conference on Machine Learning (ICML '04)}. ACM, New York, NY, USA, pp. 29-36.

\bibitem{AJ01}
A. Hyvarinen, J. Karhunen, E. Oja
\newblock {\em Independent Component Analysis,} New York: J. Wiley 2001.

\bibitem{Freedman}
D. A. Freedman,
\newblock {\em Statistical Models: Theory and Practice,} Cambridge University Press, 2005.

\bibitem{Kalman}
R. E. Kalman, ``A New Approach to Linear Filtering and Prediction Problems," {\em Transaction of the ASME--Journal of Basic Engineering}, vol. 82, no. Series D, pp. 35-45, March 1960.


\bibitem{Dempster}
A. P. Dempster, N. M. Laird, D. B. Rubin, ``Maximum Likelihood from Incomplete Data Via the EM Algorithm," {\em J. Royal Stat. Soc.} B. 39, 1, 1Ð38.

\bibitem{Singh}
M. Singh, M. Valtorta, ``Construction of Bayesian Network Structures from Data: A Brief Survey and an Efficient Algorithm," {\em International Journal of Approximate Reasoning}, vol. 12, no. 2, pp. 111Ð131, 1995. 


\bibitem{Tajer}
A. Tajer, V.V. Veeravalli, H.V. Poor, ``Outlying Sequence Detection in Large Data Sets," {\em IEEE Signal Processing Magazine}, vol. 31, no. 5, pp. 44-56, September 2014.

\end{thebibliography}
\end{document}